\documentstyle[pra,epsf,aps,twocolumn]{revtex}
\def\sbf#1{{\bf\scriptsize #1}}

\begin{document}
%%%%
%\begin{titlepage}
\title{Vortex Line Nucleation of First-Order Transition
$U(1)$-Symmetric Field Systems}
\author{H.\ Kleinert \\ Institut f\"ur Theoretische Physik\\Arnimallee 14,
D14195 Berlin, Germany
}
%%%%%%%%%%%%
%
\maketitle
{}~\\
\begin{abstract}
We show that in field systems with $U(1)$-symmetry, first-order
transitions are nucleated by vortex lines, not bubbles, thus
calling for a reinvestigation of the Kibble mechanism
for the phase transition of the early universe.
\end{abstract}
%\end{titlepage}
%\newpage
%
%
      ~\\

\noindent {\bf 1.} Since Langer's historic paper on bubble nucleation
\cite{1,2}  of first-order transition in a real scalar
field system,
field theorists have assumed this mechanism to
cause transitions
in a large variety of physical systems.
This belief was enhanced a rigorous proof
of Coleman at al.,
that the dominant
classical solutions of rotationally-invariant field
equations in instanton calculations
are bubble-like \cite{3}.
Most importantly for our very existence,
the first-order transitions in
the early universe
are
supposed
to be nucleated by
bubbles via the celebrated Kibble mechanism \cite{4,5}.

To an unbiased observer,
this assumption comes as a surprise,
since
the evolution of the early universe
is described by
a field theory
which
is
a nonabelian generalization of the
Ginzburg-Landau theory of superconductivity.
For a superconductor, however,
bubbles play no role in the phase transition.
A superconductor can have a second- or a first-order transition,
depending on the
ratio of the two length scales $ \kappa $\,=\,magnetic penetration
depth/coherence lenth \cite{tri,GFtri}.
In the second-order regime, where $ \kappa $ is large,
the transition can be
understood completely as
 a proliferation of magnetic vortex lines.
This can shown convincingly in a
lattice field theory of the system \cite{GF}.

Moreover, there is a dual description
of this theory which is a simple XY-model.
The high-temperature expansion  of the partition function
of this model
can be rewritten as a sum of closed lines
which are direct pictures of the
magnetic vortex lines in the superconductor
at low temperatures.
In this grand-canonical line ensemble
one can easily calculate the temperature of proliferation
 \cite{GF}.

In the continuous limit,
this XY-model can be transformed via functional techniques into
a $|\psi^4|$-field theory with a complex {\em disorder field\/} \cite{GF}.
In this formulation, the Feynman diagrams in the perturbation
expansion of the vacuum energy
are the direct pictures of the
magnetic vortex lines, which  proliferation
as the mass term becomes negative.

When lowering the parameter $ \kappa $ into
the regime of weak
first-order  transitions, there still exists
a generalization of the XY-model
describing this system,
which has
the same type of
high-temperature expansion
in terms of closed loops, thus
showing again that only vortex lines can be relevant
for understanding the transition  \cite{GFtri}.
Thus we must conclude that
in a superconductor and related field theories
which possess vortex lines as topological
excitation, these excitations
are also the relevant driving mechanism
of the phase transitions.

It is the purpose of this note
to make these qualitative arguments
convincing, demonstrating
 the
  superior efficiency of
vortex line over
bubble nucleation, thus casting doubts
on all studies of phase transition
based on the Kibble mechanism. \\

\noindent {\bf 2.} The generalized XY-model
 which provides us with a disorder description of
 a superconductor
on a lattice
has the
partition function
\begin{equation}
 Z = \int \frac{{\cal D}\theta}{2 \pi }  e^{ \beta \sum_{{\sbf x},i}
     \left[ \cos \nabla_i \theta +  \delta \cos 2\nabla_i \theta\right] }
\label{PF}\end{equation}
 where $\nabla_i$ are lattice gradients, and $ \beta , \delta $ model
parameters.
The phase structure of this model has been studied in detail
in the literature \cite{JK,JK2,GF}.
For $ \delta =0$  the model is known to describe the
critical behavior of superfluid helium near the $ \lambda $-transition.
The same thing is true for a small interval around zero
$ \delta \in (0.1, 0.2)$.
In addition, there  exists a regime of $ \delta $ where the transition
is of first order.
In the disordered phase,
the partition function (\ref{PF})
 can be rewritten as a sum over non-self-backtracking loops
of superflow. Under a duality transformation, these go over into
the magnetic vortex lines of the
superconductor. The parameter $ \beta $
which is the inverse temperature in the XY-model
grows with the temperature in the superconductor.
The loops of superflow can have strengths 1,2,3,\,\dots~ on the lattice.
They are dual representation of the quantized flux strengths of the magnetic
vortex lines
 in the superconductor.
In the second-order regime,
the critical properties of the model
have been shown to be the same as for a simplified model
which can contain only loops of unit strength \cite{JKV,GF}.

For a single loop,
the partition function of this simplified model
can easily be written down.
If $n$ is the length of the loop in lattice units,
we have
\begin{equation}
 Z = \sum_{n} N_n e^{ \beta _V  \varepsilon_n}
\label{Z}\end{equation}
where $ \beta _V$ is a function $ \beta , \delta $ which plays
the role of an inverse temperature
for this one-loop model,
$\varepsilon_n$ is the loop energy,
 and
 $N_n$  is the number of different loops
 of length $n$.
For large $n$,
 the energy $\varepsilon_n$ is proportional to
$n$, say $\varepsilon_n\approx \varepsilon\, n$.
The notation $ \epsilon _n$ is really an
approximation, since it  neglects
 a slight dependence on the loop shape. This, however,
 is very weak for lines which are much longer than the
length scale $n^{\rm st}$ over which the lines show stiffness.
This stiffness is a result of the
non-self-backtracking property and the fact that
if two (or more) portions of a loop
 merge into a line of strength two (or larger),
the energy of this portion is much larger than
the sum of the energies
of the constituent lines,
causing a
strong Boltzmann suppression.
Writing the
number
 $N_n$
as  $ e^{- s_n}$,
we define
 the configurational entropy $s_n$
of loops of length $n$.
Also $s_n$ grows linearly for large $n$,
say like
 $ s\, n$. As $ \beta _V $ becomes smaller
than a critical value $ \beta _{V}^c \equiv {s}/{\epsilon}$,
the free energy of the loops
\begin{equation}
   f_n = \varepsilon_n -  \beta _V^{-1} s_n
\label{}\end{equation}
goes to negative infinity for large $n$, so that
the sum over $n$ in (\ref{Z}) diverges.
The loop length
diverges
 and the loop
fills the
entire system with superflow, a characteristic feature of the
phase transition into the superfluid state.
A large energy of a loop will  always be canceled
by the configurational entropy if the temperature is sufficiently large.

\begin{figure}[tbhp]
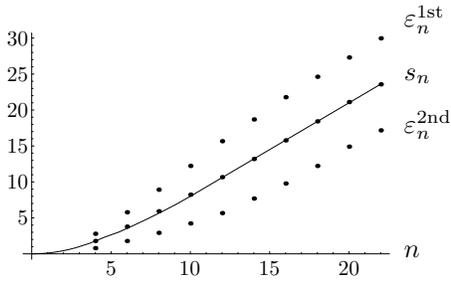

\vspace{0cm}
\input entrops2.tps ~\\
\caption[]{Dependence of the entropy
of lines of length $n$ in the partition function
(\protect\ref{Z}) on the length $n$ of the loops. The curves above and below
are possible
energies $ \varepsilon _n$ leading to
second-
or first-order
phase transitions.
 }
\label{@ent}\end{figure}

A decrease of the parameter  $ \delta $ in (\ref{Z})
brings the phase transition into the first-order regime.
In the loop picture,
this change
the $n$-dependence of the energy $ \epsilon_n$.
In the partition function (\ref{Z}),
The entropy $ s _n$
of the loops in
the small-$ \beta $ expansion of
(\ref{Z})
depends on $n$ as shown in Fig.~\ref{@ent} \cite{bc}.
After an initial rise it flattens out somewhat  around $n\approx10$,
where
it merges into the asymptotic linear behavior $ \sigma \,n$.
The energy may depend on $n$ in
different characteristic ways, also
indicated in
Fig.~\ref{@ent}.
The region  $n\approx10$ where the
linear behavior is reached
is determined by
the effective
stiffness of the vortex lines.

\begin{figure}[bh]
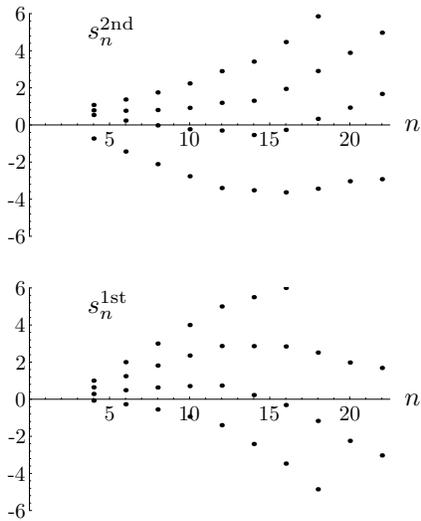

\vspace{-.3cm}~\\
\input f2nd.tps \\%~~~~~~~~~~~~~~~~~~~~~~~~~~~~~~~~~~~~
\input f1st.tps ~\\
\caption[]{Free energy as a function of loop length $n$
for second- and first-oder phase transitions at different inverse
model temperatures
$ \beta _V$.}
\label{@tr}\end{figure}

The associated free energies
$f_n$ have the typical
shapes displayed
in   Fig.~\ref{@tr}.
The left-hand plot
shows the free energy
for $ \varepsilon ^{ 2{\rm nd}}$
in an ordinary XY-model. For sufficiently large temperatures, it
 possesses a minimum  at a nonzero
value of $n$, say at $n^{\rm m}$.
This value moves  continuously from zero to infinity as
$ \beta _{V}$ is raised above
the  critical
value $ \beta _{V}^c$.
The transition is of second order.
Even before the critical value is reached, there are
loops of size $n^{\rm m}$ in the system.
  Such {\em precritical loops\/}
 are found in Monte-Carlo simulations
of the model (\ref{Z}). They are plotted
as 3D-figures in Ref.~\cite{3D}.

The free energy in the right-hand plot of Fig.~\ref{@tr}
corresponds to the energy $\varepsilon^{1{\rm st}}$, and
shows a completely different behavior.
As the critical value   $ \beta _{V}^c$
is reached, the free energy has a barrier at $n^{\rm m}$
which prevents the lines from growing infinitely long.
Thermal fluctuations have to create a loop  of length $n^{\rm m}$,
which can then expand and fill the entire system
with superflow, thereby converting the normal state of the
XY-model into
a superfluid one, or the ordered state of a superconductor
with magnetic vortex lines
into the normal state.
The size of $n^{\rm m}$ is of the order
of the length scale of stiffness  $n^{\rm st}$.

Since the superconductor
on a lattice can be represented exactly in terms
of loops,
there is no place for bubble nucleation
in such a system.
But there are also simple
energy-entropy arguments
to justify this conclusion.

\noindent {\bf 3.} Consider a possible nucleation
of the transition
by bubbles \cite{1,2,PI}.
Such bubbles can be calculated in a continuous approximation to the partition
function
(\ref{Z}) which can be derived by standard field theoretic
techniques \cite{cf,JK2}. In this approximation,
the partition function
(\ref{Z}) becomes a functional integral
over a complex disorder field $\phi(x)$
with quartic and sextic self-interactions \cite{JK2,GFtri}.
When cooling the disordered phase slightly below the transition point,
such a field theory
possesses spherically-symmetric  solutions
whose inside
contains the
ordered phase
whose energy is slightly lower than that of the  disordered phase.
Let $  \epsilon  $
be the difference in energy density and $   \sigma  $ the
surface energy density.
The total energy of the bubble is then
\begin{equation}
E ^{\rm bubble}=S_DR^{D-1} \sigma -\frac{S_D}D R^D \epsilon,
\label{@}\end{equation}
where $S_D=2\pi^{D/2}/ \Gamma (D/2)$
is the surface of a unit sphere in $D$ dimensions.
This energy is maximal at $R_c=(D-1){ \sigma }/{ \epsilon }$,
where it is equal to
\begin{equation}
E_c
%=\frac{S_D}{D}R_c^{D-1} \sigma
%=\frac{S_D}{D(D-1)}R_c^{D} \epsilon
=\frac{S_D}{D}(D-1)^{D-1}\frac{ \sigma ^D}{ \epsilon ^{D-1}}.
\label{@}\end{equation}
The important point is now that
for temperatures which lie only very little
 beyond the transition temperature,
the energy difference $ \epsilon $ between
the two
phases is
very small, corresponding a huge bubble radius and energy.
The probability of nucleating such a bubble is therefore infinitesimally small.
Only after considerable overheating (or overcooling)
does the bubble energy become small enough to
nucleate spontaneously (in the
absence of other condensation nulei such as dirt).
In freezing transition of water, the radius $r_c$ is about 50 \AA.

In a superconductor, however,
the phase transition
proceeds without overheating, and the reason for this is
 the vortex nucleation mechanism discussed above.
The energy of a critical vortex
may be estimated by imagining a planar phase boundary
 rolled up to a thin line whose radius is the
coherence length $\xi_0$
of the disorder theory. This, in turn,
is bent into a doughnut of radius $n^{\rm st}\xi_0$.
 Neglecting
the bending energy, we estimate the critical vortex energy
to be of the order
\begin{equation}
E_{\rm vort}^c\approx 2\pi \xi_0 \times
n^{\rm st}\,\xi_0 \sigma .\label{@}\end{equation}
This energy does not depend
on the energy difference $ \epsilon $ between the two phases,
so that the rate is practically independent of the
degree of overheating (or undercooling),
this being
in contrast to the energy of the critical bubble
which is extremely large slightly beyond the transition
point.

Note that in contrast to vortex nucleation,
bubble nucleation is not enhanced significantly
by the configurational entropy
of the bubble surface.
The reason is that
apart from translations,
all surface fluctuations
are massive \cite{PI},
Configurational distortions of a long vortex line,
on the other hand, require practically no energy
if taking place over length scales larger than the finite stiffness length.

\noindent {\bf 4.} The above argument
imply
that
bubble nucleation is of no relevance to
first-order phase transition
in superconductors, and
for that matter, to the first-order transitions in the early universe,
as long as the theory describing the latter allows for
line-like topological excitations. These
drive the transition with much greater efficiency
than bubbles due to their larger configurational entropy.
{}~\\~\\
{\bf Acknowledgements}\\
The author is grateful to Prof. Tom Kibble
for heated discussions at the 1997 Grenoble Meeting on
the dynamics of defects. He also thanks Dr. Alisdair Gill
for recent interactions which helped sharpening
my Grenoble arguments.
Finally, I thank Prof. Tony Leggett for an informative email on
the status  of the discussion of the AB phase
transitions in superfluid ${}^3$He.

\newpage
{}~\\[-0cm]
{\bf Figures}


\begin{thebibliography}{11}
\bibitem{1}
 J.S.\ Langer, Ann.\ Phys.\ {\bf 41}, 108 (1967).
\bibitem{2}
 For reviews see\\
 S.\ Coleman, Phys.\ Rev.\ {\bf D15}, 2929 (1977);
also in {\em The Whys of Subnuclear Physics\/}, Erice Lectures
1977, Plenum Press, 1979, ed.\ by A.~Zichichi;\\
 I.~Afflek,
Phys.\ Rev.\ Lett.\ {\bf 46}, 391 (1981);
\\
H.\ Kleinert, {\em Path Integrals in Quantum Mechanics,
Statistics, and Polymer Physics, 2nd Edition\/}, World Scientific,
Singapore, 1995\\
(www.physik.fu-berlin.de/\~{}kleinert/b3).
\bibitem{3}
S. Coleman, V. Glaser, A. Martin,
 Comm.\ Math.\ Phys.\ {\bf 58}, 211 (1978).
\bibitem{4}
T.W.B.\ Kibble, J. Phys. A {\bf 5}, 1387 (1976);
 Phys. Rep.~{\bf 67}, 183 (1980).\\
For recent discussions see \\
P. McGraw, Phys.~Rev.~ D {\bf 57}, 3317 (1998) (astro-ph/9706182).
\bibitem{5}
For a recent review see\\
W.H. Zurek, {\em Cosmological Experiments
in Condensed Matter Systems\/}.
(cond-mat/9607135).
\bibitem{tri}
 H. Kleinert,
     Lett.\ Nuovo Cimento  {\bf 35}, 405 (1982)
(www.physik.fu-berlin.de/\~{}kleinert/97).
\bibitem{GFtri}

See Chapter 13 in the textbook \cite{GF}
(www.physik.fu-berlin.de/\~{}kleinert\-/b1/gifs/v1-716.html).


\bibitem{JK}
 W.\ Janke and H.\ Kleinert,
 Phys.\ Rev.\ Lett.\ {\bf 57}, 279 (1986)
(www.physik.fu-berlin.de/\~{}kleinert/130).
\bibitem{JK2}
 W.\ Janke and H.\ Kleinert,
Nucl.\ Phys.\ B {\bf 270} [FS16], 399
  (1986)
(www.physik.fu-berlin.de/\~{}kleinert/139).
\bibitem{GF}
 H.\ Kleinert, {\em  Gauge Fields in Condensed Matter\/},
 Vol.\ I, see pp.\ 531--535
(www.physik.fu-berlin.de/\~{}kleinert/b1/gifs/\\v1-531.html).

\bibitem{bc}
Up  to $n$=12 the numbers $N_n$ are found on p.394
of Ref. \cite{GF}
(www.physik.fu-berlin.de/\~{}kleinert/b1/gifs/v1-394.html).
For $14<n<22$ the numbers have been calculated by \\
P. Butera and M. Comi, (to be published). \\
I am grateful to the authors for communicating these numbers to me prior to
publication.


\bibitem{JKV}
 W. Janke and H. Kleinert,
     Nucl.\  Phys.\ B {\bf 270}, 135 (1986)   \\
(www.physik.fu-berlin.de/\~{}kleinert/126)
\bibitem{cf}
See Chapter 5 in the textbook \cite{GF}
(www.physik.fu-berlin.de/\~{}kleinert/b1/gifs/v1-409.html).
\bibitem{3D}
 See p.~530 in the textbook \cite{GF}
(www.physik.fu-berlin.de/\~{}kleinert/b1/gifs/v1-530.html).

\bibitem{PI}
See Chapter 17 in\\
    H. Kleinert,
     {\em Path Integrals in Quantum Mechanics, Statistics and Polymer
Physics,\/}
     World Scientific, Singapore 1995,
     Second extended edition.
(www.physik.fu-berlin.de/\~{}kleinert/b3).


\end{thebibliography}
\end{document}